\documentstyle[12pt]{article}
\def\fun#1#2{\lower3.6pt\vbox{\baselineskip0pt\lineskip.9pt
  \ialign{$\mathsurround=0pt#1\hfil##\hfil$\crcr#2\crcr\sim\crcr}}}
\newskip\humongous \humongous=0pt plus 1000pt minus 1000pt

\newif\ifdtup

\def\oldreffmt#1{\rlap{[#1]} \hbox to 2\parindent{}}

\def\figfmt#1{\rlap{Figure {#1}} \hbox to 1in{}}

\def\beq{\begin{equation}}
\def\eeq{\end{equation}}

\def\bq{\begin{quote}}
\def\eq{\end{quote}}

\relax

\newcommand{\be}{\begin{equation}}
\newcommand{\ee}{\end{equation}}
\begin{document}

\begin{titlepage}
\rightline{CPT-91/P.2628}
\rightline{November 1991}

\vskip 1.5truecm
\begin{center}
{\large{\bf On polynomial solutions of differential equations\\}}
\vskip 1.5cm
\vskip 0.8cm
 {\bf A.Turbiner}\footnote
{On leave of absence from: Institute for Theoretical and Experimental Physics,
Moscow 117259, RUSSIA\\E-mail: TURBINER@CERNVM or TURBINER@VXCERN.CERN.CH}
\vskip 0.5cm
CPT, CNRS-Luminy, Marseille, F-13288, FRANCE
\\
(JMP, Nov.-Dec.92)
\end{center}
\vskip 1.2cm
\begin{center}
{\large ABSTRACT}
\end{center}
\vskip 0.5 cm
\begin{quote}

A general method of obtaining linear differential equations having
polynomial solutions is proposed. The method is based on an equivalence
of the spectral problem for an element of the universal
enveloping algebra of some Lie algebra in the "projectivized" representation
possessing an invariant subspace and the spectral problem for a certain
linear differential operator with variable coefficients. It is shown
in general that polynomial solutions of partial differential equations
occur; in the case of  Lie superalgebras there are  polynomial solutions
of some matrix differential equations, quantum algebras
give rise to polynomial solutions of finite--difference equations.
Particularly, known classical
orthogonal polynomials will appear when considering $SL(2,{\bf R})$  acting on
${\bf RP_1}$. As examples, some polynomials connected to projectivized
representations of $sl_2 ({\bf R})$, $sl_2 ({\bf R})_q$, $osp(2,2)$ and
$so_3$ are briefly discussed.

\end{quote}

\vfill

\end{titlepage}
\newpage

An importance of polynomial solutions of differential equations is
undoubtable for applications. From one side, they usually lead to closed
analytical expressions, allowing to build constructive
models of concrete physical phenomena. From other side, they can be used
as an entry for performing perturbative or iterative considerations.
In the paper we present a general method for generating
linear differential equations of various types having polynomial solutions.
The main idea is connected to exploiting so called "projectivized"
finite-dimensional representations in differential operators of Lie algebras.

Take some Lie algebra $g$ .

{\bf Definition.} Let us call a projectivized representation of the Lie
algebra $g$ the representation in a form
\be
\label{e1}
J^{\alpha}\ =\ a^{\alpha , \mu}(x) \partial _{\mu} \ + \ b^{\alpha}(x) \quad,
\quad
J^{\alpha} \in g
\ee
$\alpha =1,2,\dots ,{\it dim} g \ , x \in R^{n-1} \ , \mu = 1,2,\dots, n-1$
and \ $a^{\alpha , \mu}(x) , \ b^{\alpha}(x)$ are certain functions on
$R^{n-1}$ and generically $b^{\alpha}(x)$ are non-vanishing simultaneously
for all $\alpha 's$.

One possible way to define such a representation (1) is the following.
 One can define a linear representation of $g$ in ${\bf R}^n$ (homomorphism
$g\rightarrow gl(n,{\bf R})$). Then the expression (1) determines the
corresponding action of $g$ in the projective space ${\bf RP}^{n-1}$ in
affine coordinates. Hereafter, I will consider the restriction of the
action of $g$ on ${\bf R}^{n-1}\subset {\bf RP}^{n-1}$.

Let $g$ be a graded Lie algebra (for example, $g$ is graded, semi-simple).
Then the universal enveloping algebra $U_g$ is endowed with the gradation
corresponding to the gradation on $g$. Therefore one can give a natural
definition of gradation of a monomial in universal enveloping algebra :

{\bf Definition.} The gradation of a monomial in the generators is the sum
of the gradations of multipliers.

Considering elements of a universal enveloping algebra $U_g$ (the algebra of
all polynomials of the generators of $g$), one can define a regular element
as follows

{\bf Definition.} An element $h$ of universal enveloping algebra of a graded
Lie algebra is called {\bf regular} , if it contains at least one
 monomial in the generators  with non-negative  gradation.

 It is worth  emphasizing that a regular element $h \in
U_g$  always contains a monomial, which can be represented by  elements of the
Cartan subalgebra only (see discussion e.g. \cite{j}).

 The generators (1) act on inhomogeneous functions (or sections of a certain
bundle on ${\bf R}^{n-1}$). Let us consider a finite-dimensional
representation $M^d$ of dimension $d$ of the Lie algebra
 $g$ in the form (1), which means the existence of an invariant
finite-dimensional subspace $V^d$ in a certain functional space.

{\bf Lemma.} If a Lie algebra $g$ has a finite-dimensional representation
$M^d$ of dimension $d$, then any element $h \in U_g$ has a finite-dimensional
representation  of the same dimension $d$ .

The representations in terms of differential operators of type (1) were
introduced by S.Lie. However, they
are poorly studied yet
(history and last developments are given at  \cite{olver}). In fact, any
semi-simple Lie algebra can possess a projectivized representation (1).

Suppose that the  invariant subspace of  the finite-dimensional representation
 of $g$ in the form (1) can be parametrized by inhomogeneous polynomials
in some coordinates.

Now, one formulates the main theorem.

{\bf Theorem.} {\it If a semi-simple Lie algebra $g$ given by the
representation (1) has an invariant subspace $V^d$ parametrized by
inhomogeneous polynomials in some coordinates,  for almost any
polynomial element $h \in U_g$  the spectral problem
 \be
\label{e2}
h \varphi (x) = \varepsilon \varphi (x)
\ee
 possesses a certain number  of
eigenfunctions in the form of inhomogeneous, finite order polynomials.   If
$g$ is a graded algebra and $h$ is a
regular element containing monomials of positive gradation, in general, the
number of eigenfunctions in the form of a polynomial is equal to the
dimension $d$ of an invariant subspace $V^d$. If a regular element $h$
has no terms with the positive gradation, an infinite sequence of
eigenfunctions of (2) are represented by polynomials in some coordinates}.

This Theorem provides sufficient conditions for generating differential
equations having, at least, one polynomial solution.

Now let us consider some concrete examples.

{\bf 1}. $g=sl_2({\bf R})$.

A projectivized representation for the algebra $g=sl_2({\bf R})$ has the
form (see e.g. \cite{t})
\footnote{At the first time this representation has been described by S.Lie}
\label{e3}
\[ J^+ = x^2 \partial_x - 2 j x,\  \]
\be
 J^0 = x \partial_x - j \  ,
\ee
\[ J^- = \partial_x \ ,  \]
\noindent
where $x \in {\bf R^1}$. Here the parameter $j$ is interpreted as  the
spin (or mark) of the representation. If $2j$ is an non-negative integer ,
there will exist  an invariant sub-space of dimension $(2j+1)$ in the
space of functions on ${\bf R^1}$.  The realization of this
representation has a form of inhomogeneous polynomials in $x$
\be
\label{e4}
R = \{ 1, x, x^2,\dots , x^{2j} \} \ ,
\ee
Take, for instance, a polynomial element of $U_{sl_2({\bf R})}$ in the form
\be
\label{e5}
h_{sl(2,R)} = \sum _{n_1,n_2,n_3 \geq 0}^{n_1+n_2+n_3 \leq n}
{\cal P}_{n_1,n_2,n_3} (J^{+},J^{0},J^{-})
\ee
of non-commutative variables $J^{+},J^{0},J^{-}$ of degrees $n_1,n_2,n_3$,
respectively. Let us fix some convention about ordering of generators in
(5) to avoid the double counting due to commutation relations, e.g. in any
monomial all $J^+$ are situated at the left and $J^-$ are placed at the right.

The algebra $g=sl_2({\bf R})$ is graded and one can introduce the natural
gradation of generators (3) as $+1,0,-1$, respectively. The terms with
the positive gradation of (5) means an existence of monomials with
$n_1 > n_3$ ; zero gradation of (5) means an existence of terms  with
$n_1 = n_3$ and/or $n_2 \not = 0$ . Substituting (3) to (5) and then
considering the spectral problem (2), one arrives at the spectral
problem for the Fuchs-type operator
\be
\label{e6}
P_{2n}(x) \partial_x ^n \varphi (x) \ +\ P_{2n-1}(x) \partial_x ^{n-1}
\varphi (x) \ +\dots +\ P_{n}(x) \varphi (x) \ =\ \varepsilon \varphi (x)
\ee
where $P_k$ are polynomials with coefficients related to the coefficients
of the element (5). As a consequence of main Theorem, the spectral problem
 (6) can have up to $(2j+1)$ solutions in the form of multi-parametrical
polynomials of degree $2j$ in $x$. The number of parameters of these
polynomial solutions is defined by the number of non-trivial free parameters
 in (5). \footnote{ Counting the number of free parameters of $h$ one should
take into account all possible relations between generators (1). For instance,
 the quadratic Casimir operator in representation (3) becomes the number
\[ J^+ J^- - J^0 J^0 + J^0 = - j (j + 1), \]
It reduces effectively the number of free parameters for the second order
polynomial
element of $U_{sl_2 {\bf R}}$ .}
 It is worth noting that the Pochhammer's equation having a polynomial
solution (see e.g. \cite{kamke}, Part I, Ch.5 (22)) occurs in our scheme as
 a particular case (6) with polynomial (5) without terms of positive gradation.

 For the particular case of the second-order polynomial
 element of $U_{sl_2({\bf R})}$ ( $n=2$ in (5)), the general solution of (6)
 gives rise to the recently discovered quasi-exactly-solvable problems
\cite{t}, since always one can transform the spectral problem (6)
(using a change of variable, $x=x(z)$ and gauge transformation, $\varphi =
\psi \exp (a)$, where $a$ is some real function ) to the Sturm-Liouville
problem. All known one-dimensional quasi-exactly-solvable problems \cite{t}
 can be obtained in such a way. In general, it is 8-parametric family of the
Schroedinger operators. If one
considers the second-order elements of the form
\[
h=c_{\alpha \beta} J^{\alpha} J^{\beta} + c_{\alpha}
J^{\alpha} \quad , \ \alpha, \beta = +,-,0
\]
(assuming summation over repeating indeces) without terms of positive
 gradation
\be
\label{e7}
h_{sl_2({\bf R})} = c_{00} J^0 J^0 \ +\ c_{0-} J^0 J^- \ +\ c_{--} J^- J^- \
+\ c_{0} J^0 \ +\ c_{-} J^- \ ,
\ee
the equation (6) coincides to the hypergeometrical equation under the
condition of existance of truncated solutions. Corresponding eigenvalues are
\be
\label{e8}
{\epsilon}_n = - \ c_{00} n (n - 2j) \ + \ c_{0} n \ + \ const \ .
\ee
Let us emphasis that the number of free parameters is equal to 5, which
coincides to mysterious Ince number of free parameters of hypergeometrical
equation having truncated solutions. So, one has natural Lie-algebraic
interpretation of the Ince number.

In particular, if one takes (7) in the form
\be
\label{e9}
h_{sl_2({\bf R})} = \ J^- J^- \ -\ 2 J^0 \
\ee
the standard Hermite equation $y \prime \prime - 2xy\prime + \lambda y = 0$
  appears. The element
\be
\label{e10}
h_{sl_2({\bf R})} = \ J^0 J^- \ -\  J^0 \ + \ [(a+1) + j] J^-
\ee
 leads to the generalized Laguerre equation $xy \prime \prime
+ (a+1-x)y\prime + \lambda y = 0$
with the associated Laguerre polynomials as solutions. In turn, the element
\be
\label{e11}
h_{sl_2({\bf R})} = \ - J^0 J^0  + J^- J^- - (2j+1) J^0
\ee
leads to the Legendre equation $(1-x^2)y \prime \prime - 2xy\prime
+ \lambda y = 0$. And finally, taking \be
\label{e12}
h_{sl_2({\bf R})} = \ -J^0 J^0  + J^- J^- - (2j+1+a+b) J^0 + (b-a) J^-,
\ee
or,
\be
\label{e13}
h_{sl_2({\bf R})} = \ -J^0 J^0  + J^0 J^- - (2j+1+a+b) J^0 + (a+1+j) J^-,
\ee
one gets a symmetric form of the Jacobi equation $(1-x^2)y \prime \prime
+ [b-a-(a+b+2)x]y\prime + \lambda y = 0$, an asymmetric form $x(1-x)y
\prime \prime + [1+a-(a+b+2)x]y\prime + \lambda y = 0$ (see e.g.\cite{m}) ,
 respectively.

{\bf Remark}. Above consideration can be easily extended for the case
of several real variables taking
 $g=sl_2({\bf R})\oplus sl_2({\bf R}) \oplus \dots
\oplus sl_2({\bf R})$ $({\it d\ times})$. Particularly, if the second-order
 polynomial element of $U_g$ has the form
\[ h_{sl_2({\bf R})\oplus sl_2({\bf R}) \oplus \dots
\oplus sl_2({\bf R})} = -\sum_{i,l}^d (J_i^0 J_j^0 \ - \ \delta_{il}J_i^0
J_l^-) \ - \]
\be
\label{e14}
\ (G+2\sum_{i=1}^d j_i +1) \sum_{i=1}^d J_i^0 \  + \ \sum_{i=1}^d (\gamma_i
+ j_i) J_i^-,
\ee
one obtains the spectral problem $H\Phi=\lambda\Phi$ for the $d$-dimensional
 differential operator
\[ H = \sum _{i,l=1}^d (x_i\delta_{il} - x_i x_l)\partial_{x_i x_l} \ + \
\sum_{i=1}^d (\gamma_i - G x_i)\partial_{x_i} \]
with the $d$-dimensional Lauricella polynomials
$$
\Phi = F_A \Bigg[\matrix{M+G-1;-m_1,\ldots ,-m_d & \cr
       & (x_{1,\ldots ,}x_d) \cr
\gamma_1,\ldots , \gamma_d &\cr} \Bigg]
$$
(for definitions see e.g. \cite {kmt})  as solutions, where $m_i=2j_i$ are
 non-negative integers, $M=\sum_{i=1}^{d} m_i$ and the $\gamma_i$ are real
 positive numbers. In fact, the operator $H$ maps polynomials of maximum
 order $m_i$ in $x_i$ into polynomials of the same type. Lauricella polynomials
 play an important role as the explicit analytic solutions of the eigenvalue
problem for modified  Laplace-Beltrami operator on $d$-sphere (for further
 details see \cite{kmt}).

\noindent {\bf 2}. $g=so_3$.

 The finite-dimensional projectivized representation of $g=so_3$ is given
 by formulae (see \cite{st})
\label{e15}
\[J^1\  =\ \partial_y \ + \ y^2\partial_y\  + \ xy\partial_x\  - \ jy \]
\be
 J^2 \ = \ y\partial_x \ - \ x \partial _y
\ee
\[J^3\  =\ - \partial_x \ - \ x^2\partial_x \ - \ xy\partial_y \ +\ jx \]
where $x,y$ are the real variables, the spin $j$ is the non-negative integer
and the dimension of the invariant subspace is equal to
$(1+j)(1+j/2)$. Invariant sub-space can be parametrized by polynomials
\be
\label{e16}
R = \{ 1, x, y, x^2, xy,  y^2,\dots , x^{j}, x^{j-1}y, \dots, x y^{j-1} ,
 y^{j} \}
\ee
Taking the polynomial element of $U_{so(3)}$ in the form (5) (replacing
 formally $J^+ \to J^1,J^0 \to J^2,
J^- \to J^3$), one comes to the spectral problem
\[ P_{2n}(x) \partial_x ^n \varphi (x,y) \ +\ P_{2n-1, 1}(x,y) \partial_x
^{n-1} \partial_y \ \varphi (x,y) + \]
\be
\label{e17}
 \dots + \
P_{2n}(y)
 \partial_y ^n \varphi (x,y) + \dots + P_n(x,y)
\varphi (x,y)  \ =\ \varepsilon \varphi (x,y)
\ee
which has, in general, $(1+j)(1+j/2)$ polynomial eigenfunctions.
The examples of the polynomial solutions for the case $n=2$ are given
in \cite{st}. It is worth emphasizing
that, if $n=2$ in (5) and an element $h$ obeys the condition:
$c_{\alpha \beta}= c_{ \beta \alpha}$ and $c_{\alpha}=0$, where
$\alpha, \beta = 1,2,3$ , the spectral problem
(17) always can be reduced by a gauge transformation, $\varphi (x,y) =
\psi(x,y) \exp (a(x,y))$ to the spectral problem for the
Schroedinger-type operator in curved space \cite {st,mprst}. It gives rise
to the three parametric family of two-dimensional exactly-solvable problems
 \cite{st}.

Now let us consider a case of a Lie super-algebra.

{\bf 3}.  $g=sl(2,{\bf R|2})\oplus {\bf R}^1$.

 The generators of $g=sl(2,{\bf R}|2)\oplus {\bf R}^1$ were obtained in
 \cite{st} and have the form
\label{e18}
\[ J^+ = x^2 \partial_x - 2 j x + x \theta \partial_{\theta},  \]
\be
 J^0 = x \partial_x - j +{1 \over 2} \theta \partial _{\theta }\  ,
\ee
\[ J^- = \partial_x \ .  \]
\[ T = - j - {1 \over 2} \theta \partial_{\theta} \]
for bosonic (even) generators and
\be
\label{e19}
Q= { Q_1 \brack Q_2 } = { \partial_{\theta} \brack x\partial_{\theta}} ,
{\bar Q} = { {\bar Q}_1 \brack {\bar Q}_2 } = { x\theta\partial_x-2j\theta
 \brack -\theta\partial_x} ,
\ee
for fermionic (odd) generators, where $x$ is the real variable and $\theta$
 is the grassmanian one. Inspection of the generators shows that, if $2j$
 is non-negative integer the finite-dimensional representation exists.
 Corresponding invariant sub-space has the dimension $(4j+1)$ and can be
parametrazed as
\be
\label{e20}
R_j = (x^0,x^1,\dots,x^{2j}, x^0\theta, x^1\theta, \dots, x^{2j-1}\theta)
\ee
 Substituting (18),(19) into a polynomial element of the universal
 enveloping algebra and considering the spectral problem (2), one gets
 polynomial solutions.

 One can exploit the parametrization of
$\theta$ and $\partial_{\theta}$ by Pauli matrices $\sigma_+$ and $\sigma_-$.
Assuming the action of the generators (18),(19) on two-component spinors,
 one gets the representation the fermionic generators in the form
\be
\label{e21}
Q \ = \ { \sigma_- \brack x\sigma_-} ,
{\bar Q} \ = \ { x\sigma_+ \partial_x-2j\sigma_+ \brack
-\sigma_-\partial_x} .
\ee
The representation (21) implies that in the spectral problem (2) an
eigenfunction $\varphi(x)$ is treated  as  two-component spinor
\be
\label{e22}
\varphi(x) \ = \ { \varphi_1(x) \brack \varphi_2(x)} ,
\ee
and, finally, we come to the spectral problem for 2x2 matrix differential
 operator containing in general $(4j+1)$ polynomial solutions.

The example of the spectral problem having polynomial eigenfunctions
 is given in \cite{st}. Also in \cite{st} the corresponding
quasi-exactly-solvable spectral problem for the matrix Schroedinger operator
 is shown.

{\bf 4.} $g=sl_2({\bf R})_q$.

Using  quantum algebras, one can generate some finite-difference equations
having polynomial solutions.

Let us build a $q$-analogue of the projectivized
 representation (3) \cite{ot}
\label{e23}
\[ T^+ = x^2 D - \{ 2j \} x \]
\be
T^0 = \  x D - \hat j
\ee
\[ T^- = \ D \]
where $x$ is real variable, $\hat j \equiv {\{2j\}\{2j+1\}\over \{4j+2\}}$ ,
$\{n\} = {{1 - p^n}\over {1 - p}}$
is the so-called quantum symbol and $p$ is a number characterizing the
 deformation; $D z = 1 + p z
D$ and $D f(z) = {{f(z) - f(pz)} \over {(1 - p) z}} + f(pz) D$ is a shift or a
finite-difference operator (or so called Jackson's symbol (see \cite{e})).
The generators (23) obey the commutation relations for a quantum algebra
$g=sl_2({\bf R})_q$ corresponding to so called the second Witten's
deformation in a classification of C.Zachos \cite{z} (for details, see
\cite{ot}). In the limit $p \rightarrow 1$, (23) coincide to (3).

 A striking property of (23) consists of an existance of finite-dimensional
 representations. If $2j$ is non-negative integer, generators (23) form
a finite-dimensional representation of dimension $2j+1$ as well as for
 non-deformed case, $p=1$. Corresponding invariant sub-space can be
 presented by polynomials in $x$ of the order not higher than $2j$
 coinciding to (4).

One can build the universal enveloping algebra of $sl_2({\bf R})_q$  in the
same way as for ordinary Lie algebras
(considering all possible polynomials in generators). As well as for
non-deformed case, one can
introduce a notation of gradation. The {\bf Theorem} holds for this case
 as well. Taking the element $h_g \in U_g$ in the form (5), one comes
to the spectral problem (2) for $n$-th order
finite-difference operator \footnote {The $q$-analogue of
the spectral problem (2) is formulated as
\[h \varphi (x) = \varepsilon \varphi (px)  \]
in a little bit different way as usual(see e.g.\cite{e}). Inserting factor
$p$ in r.h.s. gives no infuence on our consideration. Therefore we neglect
 this curcumstance.}:
 \be
\label{e24}
P_{2n}(x,p) D_x ^n \varphi (x) \ +\ P_{2n-1}(x,p) D_x ^{n-1} \varphi (x) \
+\dots +\ P_{n}(x,p) \varphi (x) \ =\ \varepsilon \varphi (x)
\ee
(cf.(6)). Generically, the spectral problem (24) has $(2j+1)$ polynomial
 solutions under the conditions of the main {\bf Theorem}. In
order to reproduce the known $q$-deformed classical Hermite, Laguerre,
 Legendre and Jacobi polynomials
(for the latter, there exists the $q$-deformation of the asymmetric form
 (13) only)  (see e.g.\cite{e}),  one should consider the spectral problem
 (2) \footnote { We consider corresponding difference equations following
 the book by Exton \cite{e} : the equations
(5.6.2), (5.5.7.1), (5.7.2.1), (5.8.3), respectively.}
  for the following combinations in the  generators (23):
\[
h_{{sl(2,{\bf R})}_q} =  T^- T^- \ -\ \{ 2 \} T^0 , \]
\[
h_{{sl(2,{\bf R})}_q} =  T^0 T^- \ -\ p^{-a-1} T^0 \ + \ (p^{-a-1}\{ a+1\}
+ \hat{j}) T^- , \]
\[
h_{{sl(2,{\bf R})}_q} = - pT^0 T^0  + T^- T^- - [(2 \hat{j} -1) p
+\{ 2 \}] T^0 , \]
\[
h_{{sl(2,{\bf R})}_q} = - p^{a+b-1}T^0 T^0  + p^a T^0 T^- - \]
\be
\label{e25}
 [(2\hat{j}-1)p^{a+b-1}+\{a\}p^b +\{b\}] T^0 + (\{a\}+\hat{j} p^a) T^- \ ,
\ee
respectively.

    Due to the fact, that representations (3) and (23) for the same spins have
the same invariant functional sub-space (4), one can mix generators
(3) and (23), and construct polynomials in generators containing both (3)
and (23) as non-commutative variables. Evidently, those polynomials
possess an invariant sub-space and in general corresponding spectral
 problem has polynomial solutions. It gives
rise to the linear differential-difference equations in one variable having
polynomial solutions.

\begin{center}
{\bf Acknowledgement}
\end{center}

In closing, I am very grateful to V.I.Arnold for numerous fruitful
 discussions and the interest to the subject. I would like to thank of
  C.S.Christ, C.Duval and V.Yu.Ovsienko for the useful discussions and
 careful reading of the manuscript and the Centre de Physique Theorique
 for the hospitality extended to me, where this work has been done. Also
 I thank to W. Miller, Jr. for suggestion to consider the  Lauricella
polynomials from present Lie-algebraic point of view.

\end{document}